\documentclass[10pt,a4paper,onecolumn]{article}
\usepackage{marginnote}
\usepackage{graphicx}
\usepackage{xcolor}
\usepackage{authblk,etoolbox}
\usepackage{titlesec}
\usepackage{calc}
\usepackage{tikz}
\usepackage{hyperref}
\hypersetup{colorlinks,breaklinks=true,
            urlcolor=[rgb]{0.0, 0.5, 1.0},
            linkcolor=[rgb]{0.0, 0.5, 1.0}}
\usepackage{caption}
\usepackage{tcolorbox}
\usepackage{amssymb,amsmath}
\usepackage{ifxetex,ifluatex}
\usepackage{seqsplit}
\usepackage{xstring}

\usepackage{float}
\let\origfigure\figure
\let\endorigfigure\endfigure
\renewenvironment{figure}[1][2] {
    \expandafter\origfigure\expandafter[H]
} {
    \endorigfigure
}

\usepackage{fixltx2e} 
\usepackage[
  backend=biber,
]{biblatex}
\bibliography{paper.bib}

\let\textttOrig=\texttt
\def\texttt#1{\expandafter\textttOrig{\seqsplit{#1}}}
\renewcommand{\seqinsert}{\ifmmode
  \allowbreak
  \else\penalty6000\hspace{0pt plus 0.02em}\fi}

\makeatletter
\let\href@Orig=\href
\def\href@Urllike#1#2{\href@Orig{#1}{\begingroup
    \def\Url@String{#2}\Url@FormatString
    \endgroup}}
\def\href@Notdoi#1#2{\def\tempa{#1}\def\tempb{#2}%
  \ifx\tempa\tempb\relax\href@Urllike{#1}{#2}\else
  \href@Orig{#1}{#2}\fi}
\def\href#1#2{%
  \IfBeginWith{#1}{https://doi.org}%
  {\href@Urllike{#1}{#2}}{\href@Notdoi{#1}{#2}}}
\makeatother

\newlength{\cslhangindent}
\newlength{\csllabelwidth}
\newenvironment{CSLReferences}[3] 
 {
  \setlength{\parindent}{0pt}
  \ifodd #1 \everypar{\setlength{\hangindent}{\cslhangindent}}\ignorespaces\fi
  \ifnum #2 > 0
  \setlength{\parskip}{#2\baselineskip}
  \fi
 }%
 {}
\usepackage{calc}

\usepackage[top=3.5cm, bottom=3cm, right=1.5cm, left=1.0cm,
            headheight=2.2cm, reversemp, includemp, marginparwidth=4.5cm]{geometry}

\titleformat{\section}
  {\normalfont\sffamily\Large\bfseries}
  {}{0pt}{}
\titleformat{\subsection}
  {\normalfont\sffamily\large\bfseries}
  {}{0pt}{}
\titleformat{\subsubsection}
  {\normalfont\sffamily\bfseries}
  {}{0pt}{}
\titleformat*{\paragraph}
  {\sffamily\normalsize}

\usepackage{fancyhdr}
\makeatletter
\let\ps@plain\ps@fancy
\fancyheadoffset[L]{4.5cm}
\fancyfootoffset[L]{4.5cm}

\definecolor{linky}{rgb}{0.0, 0.5, 1.0}

\newtcolorbox{repobox}
   {colback=red, colframe=red!75!black,
     boxrule=0.5pt, arc=2pt, left=6pt, right=6pt, top=3pt, bottom=3pt}

\newcommand{\ExternalLink}{%
   \tikz[x=1.2ex, y=1.2ex, baseline=-0.05ex]{%
       \begin{scope}[x=1ex, y=1ex]
           \clip (-0.1,-0.1)
               --++ (-0, 1.2)
               --++ (0.6, 0)
               --++ (0, -0.6)
               --++ (0.6, 0)
               --++ (0, -1);
           \path[draw,
               line width = 0.5,
               rounded corners=0.5]
               (0,0) rectangle (1,1);
       \end{scope}
       \path[draw, line width = 0.5] (0.5, 0.5)
           -- (1, 1);
       \path[draw, line width = 0.5] (0.6, 1)
           -- (1, 1) -- (1, 0.6);
       }
   }

\patchcmd{\@maketitle}{center}{flushleft}{}{}
\patchcmd{\@maketitle}{center}{flushleft}{}{}
\patchcmd{\@maketitle}{\LARGE}{\LARGE\sffamily}{}{}
\def\maketitle{{%
  
  \AB@maketitle}}
\makeatletter
\renewcommand\AB@affilsepx{ \protect\Affilfont}
\renewcommand\AB@affilnote[1]{{\bfseries #1}\hspace{3pt}}
\renewcommand{\affil}[2][]%
   {\newaffiltrue\let\AB@blk@and\AB@pand
      \if\relax#1\relax\def\AB@note{\AB@thenote}\else\def\AB@note{#1}%
        \setcounter{Maxaffil}{0}\fi
        \begingroup
        \let\href=\href@Orig
        \let\texttt=\textttOrig
        \let\protect\@unexpandable@protect
        \def\thanks{\protect\thanks}\def\footnote{\protect\footnote}%
        \@temptokena=\expandafter{\AB@authors}%
        {\def\\{\protect\\\protect\Affilfont}\xdef\AB@temp{#2}}%
         \xdef\AB@authors{\the\@temptokena\AB@las\AB@au@str
         \protect\\[\affilsep]\protect\Affilfont\AB@temp}%
         \gdef\AB@las{}\gdef\AB@au@str{}%
        {\def\\{, \ignorespaces}\xdef\AB@temp{#2}}%
        \@temptokena=\expandafter{\AB@affillist}%
        \xdef\AB@affillist{\the\@temptokena \AB@affilsep
          \AB@affilnote{\AB@note}\protect\Affilfont\AB@temp}%
      \endgroup
       \let\AB@affilsep\AB@affilsepx
}
\makeatother

\renewcommand\Affilfont{\sffamily\small\mdseries}
\ifnum 0\ifxetex 1\fi\ifluatex 1\fi=0 
  \usepackage[T1]{fontenc}
  \usepackage[utf8]{inputenc}

\else 
  \ifxetex
    \usepackage{mathspec}
    \usepackage{fontspec}

  \else
    \usepackage{fontspec}
  \fi
  \defaultfontfeatures{Ligatures=TeX,Scale=MatchLowercase}

\fi
\IfFileExists{upquote.sty}{\usepackage{upquote}}{}
\IfFileExists{microtype.sty}{%
\usepackage{microtype}
\UseMicrotypeSet[protrusion]{basicmath} 
}{}

\usepackage{hyperref}
\hypersetup{unicode=true,
            pdftitle={flowMC: Normalizing-flow enhanced sampling package for probabilistic inference in Jax},
            pdfborder={0 0 0},
            breaklinks=true}
\let\addcontentslineOrig=\addcontentsline
\def\addcontentsline#1#2#3{\bgroup
  \let\texttt=\textttOrig\addcontentslineOrig{#1}{#2}{#3}\egroup}
\let\markbothOrig\markboth
\def\markboth#1#2{\bgroup
  \let\texttt=\textttOrig\markbothOrig{#1}{#2}\egroup}
\let\markrightOrig\markright
\def\markright#1{\bgroup
  \let\texttt=\textttOrig\markrightOrig{#1}\egroup}

\IfFileExists{parskip.sty}{%
\usepackage{parskip}
}{
\setlength{\parindent}{0pt}
\setlength{\parskip}{6pt plus 2pt minus 1pt}
}
\providecommand{\tightlist}{%
  \setlength{\itemsep}{0pt}\setlength{\parskip}{0pt}}
\ifx\paragraph\undefined\else
\let\oldparagraph\paragraph
\renewcommand{\paragraph}[1]{\oldparagraph{#1}\mbox{}}
\fi
\ifx\subparagraph\undefined\else
\let\oldsubparagraph\subparagraph
\renewcommand{\subparagraph}[1]{\oldsubparagraph{#1}\mbox{}}
\fi

\title{flowMC: Normalizing-flow enhanced sampling package for
probabilistic inference in Jax}

        \author[1]{Kaze W. K. Wong}
          \author[2, 3]{Marylou Gabrié}
          \author[1]{Daniel Foreman-Mackey}
    
      \affil[1]{Center for Computational Astrophysics, Flatiron
Institute, New York, NY 10010, US}
      \affil[2]{École Polytechnique, Palaiseau 91120, France}
      \affil[3]{Center for Computational Mathematics, Flatiron
Institute, New York, NY 10010, US}
  \date{\vspace{-7ex}}

\begin{document}
\maketitle

\marginpar{

  \begin{flushleft}
  \sffamily\small

  {\bfseries DOI:} \href{https://doi.org/DOI unavailable}{\color{linky}{DOI unavailable}}

  \vspace{2mm}

  {\bfseries Software}
  \begin{itemize}
    \setlength\itemsep{0em}
    \item \href{N/A}{\color{linky}{Review}} \ExternalLink
    \item \href{NO_REPOSITORY}{\color{linky}{Repository}} \ExternalLink
    \item \href{DOI unavailable}{\color{linky}{Archive}} \ExternalLink
  \end{itemize}

  \vspace{2mm}

  \par\noindent\hrulefill\par

  \vspace{2mm}

  {\bfseries Editor:} \href{https://example.com}{Pending
Editor} \ExternalLink \\
  \vspace{1mm}
    {\bfseries Reviewers:}
  \begin{itemize}
  \setlength\itemsep{0em}
    \item \href{https://github.com/Pending Reviewers}{@Pending
Reviewers}
    \end{itemize}
    \vspace{2mm}

  {\bfseries Submitted:} N/A\\
  {\bfseries Published:} N/A

  \vspace{2mm}
  {\bfseries License}\\
  Authors of papers retain copyright and release the work under a Creative Commons Attribution 4.0 International License (\href{http://creativecommons.org/licenses/by/4.0/}{\color{linky}{CC BY 4.0}}).

  \end{flushleft}
}

\hypertarget{summary}{%
\section{Summary}\label{summary}}

Across scientific fields, more and more flexible models are required to
understand increasingly complex physical processes. However the
estimation of models'parameters becomes more challenging as the
dimension of the parameter space grows. A common strategy to explore
parameter space is to sample through a Markov Chain Monte Carlo (MCMC).
Yet even MCMC methods can struggle to faithfully represent the parameter
space when only relying on local updates.

\texttt{flowMC} is a Python library for accelerated Markov Chain Monte
Carlo (MCMC) leveraging deep generative modelling. It is built on top of
the machine learning libraries \texttt{JAX} and \texttt{Flax}. At its
core, \texttt{flowMC} uses a local sampler and a learnable global
sampler in tandem to efficiently sample posterior distributions. While
multiple chains of the local sampler generate samples over the region of
interest in the target parameter space, the package uses these samples
to train a normalizing flow model, then use it to propose global jumps
across the parameter space. The \texttt{flowMC}sampler can handle
non-trivial geometry, such as multimodal distributions and distributions
with local correlations.

The key features of \texttt{flowMC} are summarized in the following
list:

\begin{itemize}
\tightlist
\item
  Since \texttt{flowMC} is built on top of \texttt{JAX}, it supports
  gradient-based samplers through automatic differentiation such as MALA
  and Hamiltonian Monte Carlo (HMC).
\item
  \texttt{flowMC} uses state-of-the-art normalizing flow models such as
  Rational-Quadratic Splines to power its global sampler. These models
  are very efficient in capturing important features within a relatively
  short training time.
\item
  Use of accelerators such as GPUs and TPUs are natively supported. The
  code also supports the use of multiple accelerators with SIMD
  parallelism.
\item
  By default, Just-in-time (JIT) compilations are used to further speed
  up the sampling process.
\item
  We provide a simple black box interface for the users who want to use
  \texttt{flowMC} by its default parameters, yet provide at the same
  time an extensive guide explaining trade-offs while tuning the sampler
  parameters.
\end{itemize}

The tight integration of all the above features makes \texttt{flowMC} a
highly performant yet simple-to-use package for statistical inference.

\hypertarget{statement-of-need}{%
\section{Statement of need}\label{statement-of-need}}

Bayesian inference requires to compute expectations with respect to a
posterior distribution on parameters \(\theta\) after collecting
observations \(\mathcal{D}\). This posterior is given by

\[
p(\theta|\mathcal{D}) = \frac{\ell(\mathcal{D}|\theta) p_0(\theta)}{Z(\mathcal{D})}  
\]

where \(\ell(\mathcal{D}|\theta)\) is the likelihood induced by the
model, \(p_0(\theta)\) the prior on the parameters and
\(Z(\mathcal{D})\) the model evidence. As soon as the dimension of
\(\theta\) exceeds 3 or 4, it is necessary to resort to a robust
sampling strategy such as a MCMC. Drastic gains in computational
efficiency can be obtained by a careful selection of the MCMC transition
kernel which can be assisted by machine learning methods and libraries.

\textbf{\emph{Gradient-based sampler}} In a high dimensional space,
sampling methods which leverage gradient information of the target
distribution are shown to be efficient by proposing new samples likely
to be accepted. \texttt{flowMC} supports gradient-based samplers such as
MALA and HMC through automatic differentiation with \texttt{Jax}. The
computational cost of obtaining a gradient in this way is often of the
same order as evaluating the target function itself, making
gradient-based samplers favorable with respect to the
efficiency/accuracy trade-off.

\textbf{\emph{Learned transition kernels with normalizing flow}}
Posterior distribution of many real-world problems have non-trivial
geometry such as multi-modality and local correlations, which could
drastically slow down the convergence of the sampler only based on
gradient information. To address this problem, \texttt{flowMC} also uses
a generative model, namely a normalizing flow (NF) (Kobyzev et al.,
2021; Papamakarios et al., 2021), that is trained to mimic the posterior
distribution and used as a proposal in Metropolis-Hastings MCMC steps.
Variant of this idea have been explored in the past few years (e.g.,
Albergo et al., 2019; Hoffman et al., 2019; Parno \& Marzouk, 2018, and
references therein). Despite the growing interest for these methods, few
accessible implementations for non-experts already exist, especially
with GPU and TPU supports. Notably, a version of the NeuTra sampler
(Hoffman et al., 2019) is available in Pyro (Bingham et al., 2019) and
the PocoMC package (Karamanis et al., 2022) implements a version of
Sequential Monte Carlo including NFs.

\texttt{flowMC} implements the method proposed by Gabrié et al. (2021).
As individual chains explore their local neighborhood through
gradient-based MCMC steps, multiple chains can be used to train the NF,
so it can learn the global landscape of the posterior distribution. In
turn, the chains can be propagated with a Metropolis-Hastings kernel
using the NF to propose globally in the parameter space. The cycle of
local sampling, NF tuning and global sampling is repeated until
obtaining chains of the desired length. The entire algorithm belongs to
the class of adaptive MCMCs (Andrieu \& Thoms, 2008) collecting
information from the chains previous steps to simultaneously improve the
transition kernel. Usual MCMC diagnostics can be applied to assess the
robustness of the inference results, thereby avoiding the common concern
of validating the NF model. If further sampling from the posterior is
necessary, the flow trained during a previous run can be reused without
further training. The mathematical detail of the method are explained in
(Gabrié et al., 2021).

\textbf{\emph{Use of Accelerator}} Modern accelerators such as GPUs and
TPUs are designed to execute dense computation in parallel. Due to the
sequential nature of MCMC, a common approach in leveraging accelerators
is to run multiple chains in parallel, then combine their results to
obtain the posterior distribution. However, a large portion of the
computation time comes from the burn-in phase for which
chain-parallelization provides no speed up. \texttt{flowMC} is built on
top of \texttt{JAX}, so that it supports the use of GPU and TPU
accelerators by default. Users can write codes in the same way as they
would do on a CPU, and the library will automatically detect the
available accelerators and use them at run time. Furthermore, the
library leverage Just-In-Time compilations to further improve the
performance of the sampler.

\textbf{\emph{Simplicity and extensibility}} We provide a black-box
interface with a few tuning parameters for users who intend to use
\texttt{flowMC} without too much customization on the sampler side. The
only inputs we require from the users are the log-posterior function and
initial position of the chains. On top of the black-box interface, the
package offers automatic tuning for the local samplers, in order to
reduce the number of hyperparameters the users have to manage.

While we provide a high-level API for most of the users, the code is
also designed to be extensible. In particular, custom local and global
sampling kernels can be integrated in the \texttt{sampler} module.

\hypertarget{acknowledgements}{%
\section{Acknowledgements}\label{acknowledgements}}

M.G. acknowledges funding from Hi!Paris.

\hypertarget{references}{%
\section*{References}\label{references}}
\addcontentsline{toc}{section}{References}

\hypertarget{refs}{}
\begin{CSLReferences}{1}{0}
\leavevmode\hypertarget{ref-Albergo2019}{}%
Albergo, M. S., Kanwar, G., \& Shanahan, P. E. (2019). {Flow-based
generative models for Markov chain Monte Carlo in lattice field theory}.
\emph{Physical Review D}, \emph{100}(3), 034515.
\url{https://doi.org/10.1103/PhysRevD.100.034515}

\leavevmode\hypertarget{ref-Andrieu2008}{}%
Andrieu, C., \& Thoms, J. (2008). {A tutorial on adaptive MCMC}.
\emph{Statistics and Computing}, \emph{18}(4), 343--373.
\url{https://doi.org/10.1007/s11222-008-9110-y}

\leavevmode\hypertarget{ref-bingham2019pyro}{}%
Bingham, E., Chen, J. P., Jankowiak, M., Obermeyer, F., Pradhan, N.,
Karaletsos, T., Singh, R., Szerlip, P. A., Horsfall, P., \& Goodman, N.
D. (2019). Pyro: Deep universal probabilistic programming. \emph{J.
Mach. Learn. Res.}, \emph{20}, 28:1--28:6.
\url{http://jmlr.org/papers/v20/18-403.html}

\leavevmode\hypertarget{ref-Gabrie2021a}{}%
Gabrié, M., Rotskoff, G. M., \& Vanden-Eijnden, E. (2021). {Efficient
Bayesian Sampling Using Normalizing Flows to Assist Markov Chain Monte
Carlo Methods}. \emph{Invertible Neural Networks, NormalizingFlows, and
Explicit Likelihood Models (ICML Workshop).}
\url{https://arxiv.org/abs/2107.08001}

\leavevmode\hypertarget{ref-Hoffman2019}{}%
Hoffman, M. D., Sountsov, P., Dillon, J. V., Langmore, I., Tran, D., \&
Vasudevan, S. (2019). {NeuTra-lizing Bad Geometry in Hamiltonian Monte
Carlo Using Neural Transport}. \emph{1st Symposium on Advances in
Approximate Bayesian Inference, 2018 1--5}.
\url{http://arxiv.org/abs/1903.03704}

\leavevmode\hypertarget{ref-Karamanis2022}{}%
Karamanis, M., Beutler, F., Peacock, J. A., Nabergoj, D., \& Seljak, U.
(2022). {Accelerating astronomical and cosmological inference with
Preconditioned Monte Carlo}. \emph{arXiv Preprint}, \emph{2207.05652}.
\url{http://arxiv.org/abs/2207.05652}

\leavevmode\hypertarget{ref-Kobyzev2021}{}%
Kobyzev, I., Prince, S. J. D., \& Brubaker, M. A. (2021). {Normalizing
Flows: An Introduction and Review of Current Methods}. \emph{IEEE
Transactions on Pattern Analysis and Machine Intelligence},
\emph{43}(11), 3964--3979.
\url{https://doi.org/10.1109/TPAMI.2020.2992934}

\leavevmode\hypertarget{ref-Papamakarios2019}{}%
Papamakarios, G., Nalisnick, E., Rezende, D. J., Mohamed, S., \&
Lakshminarayanan, B. (2021). {Normalizing Flows for Probabilistic
Modeling and Inference}. \emph{Journal of Machine Learning Research},
\emph{22}(57), 1--64. \url{https://jmlr.org/papers/v22/19-1028.html}

\leavevmode\hypertarget{ref-Parno2018}{}%
Parno, M. D., \& Marzouk, Y. M. (2018). Transport map accelerated markov
chain monte carlo. \emph{SIAM-ASA Journal on Uncertainty
Quantification}, \emph{6}, 645--682.
\url{https://doi.org/10.1137/17M1134640}

\end{CSLReferences}

\end{document}